\documentclass{sigchi}

 \toappear{%Permission to make digital or hard copies of all or part of this work for personal or classroom use is 	granted without fee provided that copies are not made or distributed for profit or commercial advantage and that copies bear this notice and the full citation on the first page. Copyrights for components of this work owned by others than ACM must be honored. Abstracting with credit is permitted. To copy otherwise, or republish, to post on servers or to redistribute to lists, requires prior specific permission and/or a fee. Request permissions from permissions@acm.org. \\
 }
\usepackage{etoolbox}
\makeatletter
\patchcmd{\maketitle}{\@copyrightspace}{}{}{}
\makeatother

\usepackage{balance}  % to better equalize the last page
\usepackage{graphics} % for EPS, load graphicx instead
\usepackage{times}    % comment if you want LaTeX's default font
\usepackage{url}      % llt: nicely formatted URLs
\usepackage{rotating} %added by poorna
\usepackage{pdflscape} %added by poorna
\usepackage{multirow}  %added by poorna
\usepackage{array, makecell}  %added by poorna
\usepackage{tabularx, booktabs} %added by poorna
\usepackage{tabu}  %added by poorna
\usepackage{boldline}
\usepackage{array,booktabs,arydshln,xcolor}

\makeatletter
\def\url@leostyle{%
  \@ifundefined{selectfont}{\def\UrlFont{\sf}}{\def\UrlFont{\small\bf\ttfamily}}}
\makeatother
\newcommand*\rot{\rotatebox{90}}

\newcommand\VRule[1][\arrayrulewidth]{\vrule width #1}

\def\pprw{8.5in}
\def\pprh{11in}

\usepackage[pdftex]{hyperref}
\hypersetup{
pdftitle={SIGCHI Conference Proceedings Format},
pdfauthor={LaTeX},
pdfkeywords={SIGCHI, proceedings, archival format},
bookmarksnumbered,
pdfstartview={FitH},
colorlinks,
citecolor=black,
filecolor=black,
linkcolor=black,
urlcolor=black,
breaklinks=true,
}

\newlength{\dummylen}
\newcommand{\NOTE}[1]{\setlength{\dummylen}{\fboxrule}\setlength{\fboxrule}{2pt}%
            \vspace{1ex}\noindent\hfill%
            \fbox{\begin{minipage}{.96\columnwidth}#1\end{minipage}}%
            \setlength{\fboxrule}{\dummylen}\hfill{}\vspace{1ex}}
            
\renewcommand{\NOTE}[1]{\ignorespaces}

\begin{document}

\title{Design Space of Programming Tools on Mobile Touchscreen Devices}

\numberofauthors{2}
\author{
  \alignauthor Poorna Talkad Sukumar\\
    \affaddr{University of Notre Dame}\\
    \affaddr{Notre Dame, IN 46556}\\
    \email{ptalkads@nd.edu}\\
 %   \affaddr{Optional phone number}
  \alignauthor Ronald Metoyer\\
    \affaddr{University of Notre Dame}\\
    \affaddr{Notre Dame, IN 46556}\\
    \email{rmetoyer@nd.edu}\\
%    \affaddr{Optional phone number}    
%  \alignauthor 3rd Author Name\\
 %   \affaddr{Affiliation}\\
 %   \affaddr{Address}\\
  %  \email{e-mail address}\\
  %  \affaddr{Optional phone number}
}

\maketitle

\begin{abstract}

While mobile touchscreen devices are ubiquitous and present opportunities for novel applications, they have seen little adoption as tools for computer programming. In this literature survey, we bring together the diverse research work on programming-related tasks supported by mobile touchscreen devices to explore the design space for applying them to programming situations.
We used the Grounded theory approach to identify themes and classify previous work. We present these themes and how each paper contributes to the theme, and we outline the remaining challenges in and opportunities for using mobile touchscreen devices in programming applications.

%We present the aspects pertaining to each theme addressed in the papers. We also provide a discussion section at the end of every theme and finally, summarize and present ideas for future work.

\end{abstract}

\keywords{
	Programming; touchscreen; source code; IDE; smartphones; tablets; 
	software development;
%	keywords should be separated by a semi-colon. \newline
%	\textcolor{red}{Optional section to be included in your final version, 
%  but strongly encouraged.}
}

\category{H.5.2.}{Information Interfaces and Presentation (e.g. HCI)}{User Interfaces}
\category{D.1}{Programming Techniques}{}
\category{D.2.6}{Software Engineering}{Programming Environments}

%See: \url{http://www.acm.org/about/class/1998/}
%for more information and the full list of ACM classifiers
%and descriptors. \newline
%\textcolor{red}{Optional section to be included in your final version, 
%but strongly encouraged. On the submission page only the classifiers’ 
%letter-number combination will need to be entered.}

\section{Introduction}

Mobile touchscreen devices are now widely used throughout society. 
Not only do they enable new applications and bring computing to new venues, %such as mobile commerce and healthcare \cite{dinh2013survey, lane2010survey}, 
but they are also increasingly replacing or being used in combination with desktop PCs and laptops for many applications. 
%\NOTE{A citation for the above statement would be great if you can find one}
However, one area that has been relatively slow in adapting to these devices is software development. This is understandable considering the drawbacks of mobile devices. The absence of a physical keyboard, small screen space and limited processing capabilities are some of the barriers preventing them from becoming usable platforms for programming \cite{Almusaly15, Raab13, seifert11}. 

Traditional programming environments or Integrated Development Environments (IDEs) used on desktop PCs and laptops typically have been designed for use with a mouse and a physical keyboard \cite{Raab13}. Porting these environments as they are to mobile devices would be impractical and result in many usability issues \cite{Biegel14, Raab13}. Hence designing for mobile devices requires keeping in mind their drawbacks and interaction style. This raises questions regarding how and what kinds of usable software development can be enabled on mobile devices by leveraging their unique capabilities and in what contexts can they be useful.

Although it is unlikely that professional developers will use mobile devices as \textit{primary} programming devices, enabling usable programming on mobile devices can potentially benefit students in educational settings, end-user programmers, and collaborative teams, just to name a few. Mobile devices are being considered for use in classrooms for computer science programming courses; therefore, making programming environments usable on these devices could benefit these efforts \cite{Karavirta12, Tillmann12}. Additionally, many visual programming languages and languages for novice programmers can be implemented on mobile touch devices, making them accessible to beginners of all ages \cite{Kelleher05, McDirmid11}. %include few other popular visual programming papers

Mobile devices can also be used as peripheral devices in professional programming environments to facilitate certain programming tasks \cite{Bragdon11, Parnin10}. Programming can also be a collaborative process and using mobile devices such as phones and tablets in combination with desktops or even by themselves can promote collaborative and co-located programming tasks such as code review, problem solving and pair programming \cite{Bragdon11, Lichtschlag10, Parnin10}. They can also give programmers a sense of uninterrupted programming when they are in transit or are away from their desktop PCs \cite{Feldman15}. Another important strength of mobile devices is that they enable multimodal input such as speech, pen, and image input. Pen input, in particular, can be incorporated in the above mentioned scenarios \cite{Lichtschlag10, Parnin10}.

Although we can see that the notion of enabling usable programming on mobile devices is interesting and potentially useful, the relatively slow and rather skeptical adoption of these devices for programming so far tells us that it is also nontrivial. While previous work done in this area is somewhat varied and provides different perspectives on the topic, we should keep in mind that this is a problem with a number of variables and numerous possibilities and that each of the previous articles has focused on one particular facet of the problem and has designed, for example, for a particular device type, context, and set of programming tasks. 

Our goal in writing this survey article is to coalesce the research work done so far in reference to using mobile touchscreen devices for programming purposes to identify the design space and challenges in bringing programming tasks to them.
Given the variety in the past work, we aim to provide a coherent review by categorizing related work under themes. % and discussing the content of the articles under sub-themes relevant to each theme. 
We hope to provide a clear big picture as well as some useful low-level details covered in existing work and make apparent the possibilities explorable on these devices for programming. 
This article will be useful to anyone who is doing research in related areas or intends to use these devices for programming or develop software to support programming on them.
%on these devices or is simply interested in the topic. 

\subsection{Scope of the Survey}
%We first conduct a survey of existing work in the area.  
For the purposes of this survey, we only include mobile devices with touchscreens enabling direct touch and/or pen input. We mainly cover programming support enabled on smartphones and tablets. However, we also discuss the programming functionality designed for tablet PCs or laptops with touchscreen displays only if the functionality could also be used exclusively on tablets as well without requiring the use of a physical keyboard.
%, e.g., gestures formulated for programming tasks \cite{Biegel14}. 
We do not discuss the work done on the interactive tablet devices of the 60s and 70s \cite{Anderson72, Davis64} in order to keep the focus on more recent work and the significantly evolved tablet devices of today.
%also since the tablet devices used today are much different from the ones described in those papers.

%We do not include the work done on the interactive tablet devices of the 60s and 70s since they incorporated indirect interaction by using a CRT display to output what was entered on the tablet with a pen/stylus \cite{Davis64, Anderson72}.

We include papers that discuss the use of mobile devices, either as primary devices or in combination with desktops or other devices, to enable some aspect of programming. We do not include work done on the broader non-programming tasks involved in software development. %, such as requirements analysis or high-level design of a software program prior to the actual coding process. %We also primarily focus on the interface design aspects of the programming functionality enabled on the devices and discuss the execution or build aspects of programs only if they are addressed in the papers.
%all the work discussed includes some kind of interaction with the source code or program.

The articles surveyed are summarized in Table 1. 
There are a total of 23 articles that have been published between 2006 and 2016 with the bulk of them (20/23) published between 2010 and 2015. Most of the articles have been published at ACM or IEEE conferences on HCI and/or software engineering. % has been completed after the significant technology shift in the late 2000s when touchscreen mobile devices started to become ubiquitous. 
Our search terms included ``programming + phone", ``programming + touchscreen", ``source code + touchscreen", ``programming + tablet", and ``software development + touchscreen". 
We found other articles using forward and backward citation tracking. %Many of the articles discussed in this survey were published at HCI and software engineering conferences. %, such as CHI, MobileHCI, and ICSE.

\subsection{Taxonomy}

The key research questions answered by the survey are as follows: 

\begin{itemize} %[noitemsep, nolistsep]
    \item What are the main themes in which the ideas or implementations presented in the papers can be categorized based on their primary focus and contributions? 
    \item In each of the above themes, what kinds of programming functionality have been enabled or addressed on the mobile touchscreen devices? 
    \item In what ways have the papers leveraged the interaction style and features of mobile touchscreen devices to enable or address the above programming functionality? 
    \item How have the implementations been received by users?
\end{itemize}
We used the Grounded theory approach to literature review  \cite{wolfswinkel13} to identify the themes. %We read the selected articles at random and underlined excerpts that seemed relevant to the review's scope and the above research questions. We gathered concepts in the highlighted excerpts, identified a set of themes, discovered their interrelations and further refined them employing the principles of open coding, axial coding, and selective coding.   % and sub-themes relevant to each theme.
%We present `categories' under each theme to group related papers. These categories, however, are not  orthogonal and there is some overlap among them.  We place a paper in a category based on its primary goal with respect to the theme discussed.  While this goal may be clearly articulated in some cases, in other cases, it required our judgement as authors of the survey.
We present the main themes identified in the papers in the following sections and in each section, discuss the papers belonging to the corresponding theme with respect to the remaining research questions before concluding with a discussion. %areas for further work. 
A paper discussed under a theme may contain aspects pertaining to a different theme and in such cases, we briefly mention those aspects of the paper in the latter theme.
%We generally address the second and third research questions, i.e. programming functionality and mobile device interaction, mentioned above collectively because they are interconnected topics. We also address broader topics such as the contexts in which the work presented are intended to be used or the motivations for the work done if they are clearly articulated in the papers. %While we briefly discuss straightforward or recurrent topics encountered in the papers (for example, support for code annotation or Visual programming-based applications), we elaborate on those papers containing more significant and novel contributions which can provide readers with useful and diverse ideas and promote research in this area. 
%We especially elaborate on the novel contributions of papers, as opposed to more straightforward or recurrent topics, e.g., support for code annotation or visual programming.
It should be noted that many of the papers discussed in this survey present design sketches or prototypes and don't include evaluation studies. %For the papers that include evaluation studies, we mention the key findings and readers are referred to the respective papers to find the details. 
A summary of the survey is presented in Table 1. % (included as supplementary material).

\vspace{3pt}

\section{Theme 1: For smartphones, on smartphones}

In this section, we discuss work that involves support for developing applications (apps for short) on mobile phones which can also be deployed on the phones. While \textit{TouchDevelop} \cite{Tillmann11} and \textit{Mobidev} \cite{seifert11} implement complete environments for developing apps, \textit{GROPG} \cite{Nguyen13} presents a complementary feature enabling debugging of mobile apps using only smartphones.   

%We begin by presenting the motivations for the work done in these papers and then discuss the programming features specific to mobile-app development enabled and smartphone-interactions leveraged to enable the features, key results of user studies followed by a discussion section. 

%\subsection{Motivations}

%\subsection{Mobile app-development support and Interaction designs}

\subsection{Novel programming language with features amenable to smartphone interaction}

\textit{TouchDevelop} is a new programming environment tailor-made for developing and deploying mobile apps using only smartphones and by using the content, sensors and features present on the phones \cite{Tillmann11}. %They envision this environment being used to teach programming in classrooms and enabling students to practice programming anywhere on their mobile devices. 
%A variety of programs can be written using \textit{TouchDevelop}, ranging from finding the factorial of a number to using the sensors on the phone.
Although classrooms are perceived as the main context for \textit{TouchDevelop}'s intended use \cite{Tillmann12, tillmann12_2}, later studies show that it is largely used by end-users who are mostly inexperienced programmers \cite{athreya12, li13}.

%The \textit{TouchDevelop} programming environment enables creation of mobile applications on \textit{Windows} phones using the content, sensors and features present on the phones \cite{Tillmann11}. 
%To this end, a new programming language, 

%The \textit{TouchDevelop} language was designed keeping in mind the input and support for program compilation and execution on smartphones. %At the highest level of \textit{TouchDevelop}, are `scripts' that are composed of `actions'. An `action' serves as a function to achieve a particular goal and is made up of statements that in turn contain `expressions'. 
%Some of the features that \textit{TouchDevelop} offers include built-in primitives and types to access the phone's content, sensors, and web, a structured editor, syntax-coloring, type-checking in support of error messages, suggestions for autocompletion, help, an interpreter which also handles `tombstoning' (the concept of reviving a saved application after an interrupt), and an output console (called the `wall'). 
Although some of \textit{TouchDevelop}'s features were designed specifically to promote ease of use on phones, such as being statically typed to enable autocompletion, and with deliberate limitations, such as no support for user-defined types, it is still designed to contain many features common to imperative, object-oriented, and functional programming languages \cite{Tillmann11}. This helps users transfer their skills from these paradigms to \textit{TouchDevelop}. 
The environment is also designed for use on a cloud-connected device to enable sharing of scripts. 
%Various sample scripts are also included with \textit{TouchDevelop}. 

%The UI design of \textit{TouchDevelop} has been governed by the interaction style of smartphones \cite{Tillmann11}. %The aforementioned scripts, actions and expressions are displayed and edited on multiple screens or pages. 
The UI elements, consisting of code blocks, menu options and custom keypads, are designed to be easily
selectable using finger taps. % and they mainly consist of selectable statements, buttons for editing operations, and a list of statement options at the action-level, and custom keypads with a `calculator-view' to enter `tokens' at the expression-level.
Programming using \textit{TouchDevelop} requires the use of
the regular on-screen keyboard only to enter a string literal or
to rename a variable. %Some of the touch gestures commonly used on smartphones, such as pinching or tapping to zoom in before selection, and tapping and holding for the cursor to appear, were not incorporated in \textit{TouchDevelop} because these gestures either caused the viewing area to change often or took too long to execute.

A study involving the analysis of 209 \textit{TouchDevelop} scripts \cite{athreya12} found that about one third of the scripts had no functional purpose and had a low code reuse ratio (5\%) \cite{athreya12}. A large-scale longitudinal study involving 17,322 \textit{TouchDevelop} scripts and 4,275 end-users \cite{li13} found that there was high code reuse ratio (58\%) and additionally, that most of the scripts were small in size
%(73\% were less than 100 lines of code) 
\cite{li13}. Most users were novices and were active initially but later left or became less active \cite{li13}.

%Regarding users' expertise and behaviors, it was found that most of them were novices (78\%) and most of them were active initially and later left (68.3\%) or became less active (22.1\%) \cite{li13}. %have been conducted to study \textit{TouchDevelop}'s usage characteristics and users' behaviors. %While the earlier smaller study found that about one third of the scripts had no functional purpose and had a low code reuse rate (5\%) \cite{athreya12}, the longitudinal study found that there was high code reuse ratio (58\%) and additionally, that most of the scripts were small in size (73\% were less than 100 lines of code) and made heavy use of external methods \cite{li13}. Regarding users' expertise and behaviors, it was found that most of them were novices (78\%) and most of them were active initially and later left (68.3\%) or became less active (22.1\%) \cite{li13}.

\subsection{Use of a graphical editor and camera feature to facilitate app-development}

%\textit{Mobidev} is an application development environment intended for use on smartphones \cite{seifert11}. 
%\textit{Mobidev} enables development of web-based applications with simple GUIs such as survey collection apps which can also be deployed on smartphones.
%Visual programming is also used in building end-user programming applications \cite{Nardi93}. 
%\textit{Mobidev} is an end-user development environment built for creating web-based mobile applications using only smartphones \cite{seifert11}. %It provides three ways to develop apps, one of which enables the creation of the front end of the app using a visual programming-based graphical editor called the `UIBuilder'. 
Motivated by the widespread availability of smartphones in emerging countries where desktops and laptops are relatively less accessible, Seifert et al. present \textit{Mobidev} which is a development environment built for creating
web-based mobile apps using only smartphones \cite{seifert11}.
%It provides three ways to develop apps, two of which require very little code to be typed and hence allow faster development.
%The motivation behind Mobidev is the widespread availability of smartphones in emerging countries where desktops and laptops are relatively less accessible. 
Hence, enabling the creation of apps
%, e.g., a survey collection app, 
with basic GUI elements on smartphones could
potentially benefit many kinds of users including programmers,
students, and end-users in many contexts
including classrooms and workplaces.

%\textit{MobiDev} is an end-user environment developed to enable the creation of web-based GUI apps on smartphones \cite{seifert11}. 
\textit{MobiDev} is a hybrid environment combining visual and text-based programming. The UI of the apps can be created by either using a graphical editor called the \textit{UIBuilder} or by scanning a UI sketch of the app using \textit{SketchBuilder} \cite{seifert11}. %to process an image of the UI sketch of the app to build the front-end, or type the entire code for both the front-end and program logic \cite{seifert11}. 
The UI elements or widgets supported include text fields, drop-down boxes, buttons, radio buttons and checkboxes. %\textit{Puzzle} \cite{danado12} is another example VP-based end-user programming application built for use on mobile touchscreen devices.

%Two of the three interaction modes that \textit{Mobidev} provides for app-creation, namely, 
%The \textit{UIBuilder} and \textit{SketchBuilder} enable faster development using the touchscreen and camera features of smartphones, respectively. 
\textit{UIBuilder} employs touch gestures to select widgets from menus and place them on the screen using drag-and-drop. \textit{SketchBuilder} uses a camera image of a paper UI sketch as input and performs some basic image processing to identify the UI elements and workflow. The recognized UI elements can then be edited with the \textit{UIBuilder}.

The evaluation of the \textit{UIBuilder} and \textit{SketchBuilder} of the \textit{Mobidev} environment found that users spent a significant amount of time typing the back-end of the apps. Users spent more time drawing sketches and using the SketchBuilder than the UIBuilder and in general, preferred using the SketchBuilder stating that it was ``fun'' to work with.

\subsection{Code folding and transparent overlays for compact representation}

Traditionally, debugging of mobile phone apps is either done by running a desktop debugging application on an emulator, or by connecting a mobile device to a desktop or laptop computer. GROPG (GRaphical On-Phone debuGger) is a graphical debugging tool that provides the features of desktop debuggers \emph{on} \textit{Android} phones \cite{Nguyen13}.

%\textit{GROPG} (GRaphical On-Phone debuGger) also targets programmers of mobile phone applications but focuses on debugging \cite{Nguyen13}. Traditional debugging of mobile phone apps is done either by running a desktop debugging application on an emulator or by using the USB debugging feature by connecting a desktop or laptop computer to a mobile device. \textit{GROPG} is a graphical debugging tool built to make debugging apps possible using only smartphones by providing the features available on desktop debugging applications. 

%\textit{GROPG} is a graphical debugging application built for debugging \textit{Android} apps on \textit{Android} phones \cite{Nguyen13}. 
%The implementation techniques of \textit{GROPG} \cite{Nguyen13} can also be used for building debugging tools for \textit{iOS} and \textit{Windows} phone apps on their respective mobile devices. 
\textit{GROPG} provides features such as viewing the debuggee's source code in context, setting and editing breakpoints, inspecting and changing the debuggee's memory values, viewing current threads and runtime stacks, and a user interface providing simultaneous views of the debuggee and debugger. %\textit{GROPG} acts like a graphical user-interface of the (command-line) Java Debugger (JDB). While typical debugging is not built in \textit{TouchDevelop} \cite{Tillmann11}, \textit{GROPG} can be compared with Android's \textit{DroidDebugger} \footnote{\url{https://play.google.com/store/apps/details?id=net.sf.droiddebugger&hl=en}} which has a heavy text-based interface and lacks the graphical, interactive UI of \textit{GROPG}.

In order to provide the debugging features that are available
on desktop applications on space-constrained smartphones, \textit{GROPG} implements a graphical and interactive user interface
with expandable fields i.e. code folding, and a transparency-adjustable debugger pane on top of the debuggee application to free the users
from switching screens to interact with the debuggee and debugger \cite{Nguyen13}. 
%When the execution of the debuggee program is paused at a breakpoint, the transparent debugger pane is displayed enabling debugging actions to be performed. These actions are then saved, the pane is removed from view and it reappears at the next pause. 
The user interface is accessed using tap and multitouch.

%While typical debugging is not built in \textit{TouchDevelop} \cite{Tillmann11}, \textit{GROPG} can be compared with Android's \textit{DroidDebugger} \footnote{\url{https://play.google.com/store/apps/details?id=net.sf.droiddebugger&hl=en}} which has a heavy text-based interface and lacks the graphical, interactive UI of \textit{GROPG}.
In a preliminary study, \textit{GROPG} was compared with Android's \textit{DroidDebugger} \footnote{\url{https://play.google.com/store/apps/details?id=net.sf.droiddebugger&hl=en}} which has a heavy text-based interface and lacks the graphical, interactive UI of \textit{GROPG}.
%for debugging three sample applications. It was found it 
Users took less time and fewer number of steps to perform the debugging tasks using \textit{GROPG}. %\textit{GROPG}, however, had a higher memory overhead ($\sim$25MB) but could easily be accommodated given a smartphone's typical RAM of 2GB.

\subsection{Discussion}
%RAM Not sure how much this discussion section adds.

%There are certain advantages that app development (including games) on mobile devices can provide compared with conventional mobile app development. 
Studies have shown that one of the main challenges that mobile app developers face is the complexity involved in testing their apps due to emulators lacking necessary features of mobile devices including the various sensors, gesture-support, and network facilities \cite{joorabchi2013real, wasserman2010software}. Therefore, enabling a wide range of testing capabilities on mobile devices, similar to GROPG \cite{Nguyen13}, can be useful. 
Additionally, mobile platforms are becoming increasingly different from one another limiting the creation of platform-independent apps %and developers are having to write code from scratch for the same apps to be used on different platforms 
\cite{joorabchi2013real, wasserman2010software}. This problem can be countered by enabling customization of apps on individual phones to use the respective phone's resources and for deployment on that phone itself in a convenient manner.
%\emph{dedicated} development environments for use \emph{on} mobile devices such as \textit{TouchDevelop}\cite{Tillmann11} can enable creation of apps using a particular device's sensors and features, for use in a variety of contexts, and for deployment on that device itself in a convenient manner. 

One of the challenges in supporting app development on phones is sustaining the interests of users and preventing decreased usage after the initial novelty period, a trend observed in the usage of \textit{TouchDevelop} \cite{li13}. While it is understandable that mobile devices can support creation of apps up to only a certain level of complexity, enabling an expansive set of capabilities and opportunities for creatively building apps using several of the device's features may help counter this challenge.
%Past studies on general mobile developers' behaviors \cite{joorabchi2013real} and \textit{TouchDevelop} \cite{li13} have reported a high code reuse ratio. Hence making a variety of scripts and templates available for app development on mobile devices can also reduce the amount of typing needed to enter code from scratch. 

Effective use of screen real estate is another challenge in mobile app development. %and developers are often concerned about testing their applications on devices with varying screen sizes \cite{joorabchi2013real}. 
When developing apps on the phone itself, the front-end or UI of the apps can be designed by applying Visual Programming principles (discussed later) to available UI components, similar to the \textit{UIBuilder} of \textit{Mobidev} \cite{seifert11}, so that developers can directly visualize and manipulate the UI of their apps.  

While \textit{TouchDevelop} \cite{Tillmann11} is a new language designed for smartphone usage, it is challenging to enable the use of conventional ``text-heavy" programming languages, such as \textit{Java} and \textit{JavaScript}, for app development on mobile devices. This may require adapting the representation and input mechanisms of these languages for mobile device usage. We present a few ideas for such adaptations in the next section.
%, conventional programming languages, such as \textit{Java} and \textit{JavaScript}, can also be used for app development on mobile devices. However, it will be necessary to adapt both the representation and input mechanisms of these languages for mobile device usage. We present a few ideas in the next section involving leveraging the syntax and semantics of programming languages to make them suitable for use on mobile devices.

\section{Theme 2: Programming-language-driven interface designs}

To overcome the drawbacks of mobile devices, such as small screen space and difficulty using the virtual keyboard, certain papers have turned to finding programming language characteristics that are suitable for use on mobile devices. These characteristics could refer to a language's programming paradigm, its syntax or other attributes that make it possible for it to be represented compactly, input in various ways, or enable features such as autocompletion \cite{Almusaly15, Feldman15, Hesenius12, mbogo2016design}. This also includes the language design of \textit{TouchDevelop} \cite{Tillmann11} which was developed specifically for smartphone interaction. 
In this section, we discuss these resulting programming environment designs \cite{Feldman15, Hesenius12, mbogo2016design} and tool implementation \cite{Almusaly15}.

\subsection{Concatenative Programming languages are tablet-friendly}

Hesenius et al. present a sketch for a development environment to enable programming in \textit{Factor} on tablets \cite{Hesenius12}. They opine that productive programming environments for concatenative and stack-based programming languages, such as \textit{Forth} and \textit{Factor}, can be developed on space-constrained devices since they have a minimalistic and compact representation unlike imperative programming languages which are text heavy. %Additionally, traditional IDEs for mainstream languages like \textit{Java} and \textit{C} are feature-rich and also have separate design and runtime environments. Therefore, productive programming environments for concatenative languages can be developed on space-constrained devices by incorporating relatively fewer features. 
Additionally, they can also realize the languages' support for 
interactive development wherein the design environment has access to the runtime environment and vice versa, allowing code to be modified at runtime. Their prototype sketch shows the tablet screen divided into \textit{Factor}-specific tools including a program editor, tool for code-navigation, and stack visualizations for the code \cite{Hesenius12}. %Other features include syntax coloring, search option, and support for interactive development (any changes made to the code will be automatically reflected in the stack visualizations).

\subsection{Using speech templates to input Java code}

Feldman et al. present ideas for implementing a programming environment, \textit{Deverywhere}, for existing languages such as \textit{Java} and \textit{JavaScript} to be used on smartphones and tablets \cite{Feldman15}. Voice and touch input are proposed for program entry, editing and navigation in \textit{Deverywhere}. %Inputting source code efficiently using speech is non-trivial but the authors state that it is possible 
This can be achieved by incorporating the programming language's syntax and semantics, and context-sensitive capabilities in the speech recognition tool \cite{begel05, Desilets06}. They list a few guidelines for creating speech templates, e.g., saying ``for i from zero to n" to input the \textit{Java} code, \texttt{for (int i = 0; i<n; i++)} on the mobile device and these templates can also be customized to suit the styles of different programmers.

%\textit{Deverywhere} is intended to be used on smartphones and tablets and to e 
Enabling programming mainly using voice input %in \textit{Deverywhere}
 reduces or eliminates the need for using the soft
keyboard \cite{Feldman15}. \textit{Deverywhere} also focuses on providing alternative and compact representations for displaying the code on mobile devices. % using ideas from literature on abstractions of code and multiple views of software \cite{Davis10, nunez12, simonyi06}. %also focuses on displaying the code in a compact form to use the limited screen space economically. 
The design suggestions for code representation include using different colors, fonts, and symbols to distinguish between the program elements, vertical text boxes to show names and extents of functions, background colors to denote scopes, and circular watermarks to denote loops. The examples shown also use bold horizontal lines as delimiters and omit some elements such as braces, types, and certain keywords, such as \texttt{then, else, class, return}, by making their connotations apparent in the code graphically. %Additionally, the speech templates used for code input can also be used for code representation.

%In \textit{RefactorPad} \cite{Raab13, raab16}, Raab et al. present a set of common editing and refactoring tasks and their corresponding gestures intended to be used as part of a standalone development environment on touchscreen devices. %obtained as a result of performing a `guessability study' similar to the one described in the paper by Wobbrock et al. \cite{wobbrock09}. The gesture set is intended to be used as part of programming environments designed for tablet devices. 
%The non-programming-language-specific list of editing and refactoring tasks was compiled by studying various editors including \textit{Eclipse}, \textit{Sublime Text}, \textit{Visual Studio} and \textit{Xcode}.
%\NOTE{And include....list the editing/refactoring tasks supported here}

\subsection{Keyboard extension to facilitate entering Java code}

Almusaly and Metoyer focus on text input and present a block-based keyboard extension for \textit{Java} program-entry on tablet devices by incorporating the language grammar
%, i.e. syntax-directed, 
and the commonly used program constructs found by analyzing \textit{Java} programs \cite{Almusaly15}. The main motivation behind this keyboard extension is the difficulty of typing on virtual keyboards, especially for programs containing many numbers, special characters, and symbols.% \cite{Almusaly15}.
One of the main design decisions made in the keyboard extension is having the keys represent frequently used program constructs rather than characters which enables faster source code input. %Keys are arranged and color-coded based on the construct types, in accordance with Gestalt principles \cite{peterson13}. 

%The keyboard extension utilizes the hierarchical structure of \textit{Java} programs by including a dynamically changing row that presents options based on the key previously pressed. Following a key-press, the template, i.e. boilerplate, corresponding to the program construct is inserted, or the user is provided with dynamic options as mentioned above, or the user is prompted to fill in details using the virtual keyboard of the device, thereby promoting correct syntax at the statement level. This also relieves the users from remembering the syntax for those constructs as well as having to hunt and peck the subsequent keys. 

%The templates for the program constructs are automatically inserted when the corresponding keys are selected, and options for subsequent code entries are presented on the keyboard, thereby relieving the users from remembering the syntax for those constructs as well as having to `hunt and peck' the subsequent keys. 

This syntax-directed keyboard extension %developed by Almusaly and Metoyer \cite{Almusaly15} 
was compared with the standard virtual keyboard with respect to the input of 2 \textit{Java} programs using 27 participants. The participants made fewer errors and used fewer keystrokes per character using the syntax-directed keyboard extension and also found the keyboard extension to be mentally less demanding than the standard virtual keyboard.

\subsection{Decomposition of Java programs to facilitate learning}

Motivated by the lack of desktops outside classroom environments, Mbogo et al. have developed a mobile application to support university students in learning \textit{Java} programming  \cite{mbogo2016design}. The main interface of the application presents the structural components of typical \textit{Java} programs and students edit a program component-wise. 
Static scaffolding learning techniques are utilized in the application wherein component-wise editing is always enforced irrespective of the level of learning attained by the student. For example, beginners are forced to complete components in a certain order while there is no order restriction for more advanced learners but all learners are required to edit the programs component-wise.
Tapping a component opens an editor tab where the students edit the respective component and another tab enables them to see the whole program or get an overview of how the components are interconnected. 

The application was evaluated using 64 students across three African universities and was deployed on an \textit{Android} phone with a screen size of 2.8 inches. Overall, the application was deemed useful for learning and the scaffolding technique of constructing a program one part at a time made it easier to program on the small screens.

\subsection{Discussion}

The key challenge here is finding ways to leverage characteristics of programming languages and tasks to support their use on mobile devices and integrating mobile device interaction mechanisms into their usage. The work discussed provides us with a few approaches. 
For example, \textit{Deverywhere} \cite{Feldman15} borrows from literature on inputting programs by voice that were originally devised for programmers suffering from Repetitive Strain Injury (RSI) and on multiple views or abstractions of code to design their interface.
Similarly, we can borrow from existing principles in programming language designs and software engineering. % to enable usable programming on mobile devices. 
Practices advocating code reuse, e.g., software design pattern, methods for code abstractions including automated refactoring, programming paradigms that enable decomposition of bigger programs into smaller manageable blocks, such as modular programming, structured programming and object-oriented programming, and techniques for concise visual presentation of programs such as code folding can all be used to make programming interfaces mobile-device-friendly and less text heavy.

%Programs can be represented in compact forms \cite{Feldman15, mbogo2016design}, or code can be entered, manipulated, and navigated using the interaction and input mechanisms afforded by the mobile devices \cite{Tillmann11, Almusaly15, Feldman15}, or programming environments requiring limited basic functionality can be implemented on these devices \cite{Hesenius12}. These approaches can also be combined.

The aforementioned techniques or programming language designs can be combined with the interaction and input mechanisms of mobile devices to enable more convenient code entry, manipulation, and navigation. %Conventional source code representations can be rethought in the light of mobile UI design patterns \cite{neil2014mobile}. 
For example, instead of plain text, UI blocks, can be used for both code entry and representation. This is observed in the block-based code used in both \textit{TouchDevelop} \cite{Tillmann11} and the ``syntax-directed" keyboard extension \cite{Almusaly15} that are easily selectable using finger taps. Pen and touch gestures can be devised for code entry, navigation, display, and for programming tasks such as refactoring (discussed later).  

%Similar to speech-based programming, source code can also be input by handwriting using a stylus on the mobile device or taking a snapshot of a piece of handwritten code, and by incorporating the programming language's syntax and semantics to the recognition \cite{gonzalez12}.

\section{Theme 3: Visual-Programming-based applications}

In this section, we present general Visual Programming (VP) concepts and briefly list examples of VP-based environments implemented on mobile touchscreen devices. We focus on the VP concepts because there is a strong association between VP and touchscreen interaction \cite{ Essl10, hackett12, Knaus08}. %and aspects of VP can be observed even in the syntax-enforcing implementations of text-based programming on mobile devices such as `TouchDevelop' \cite{Tillmann11} and the `syntax-directed' keyboard \cite{Almusaly15}.  
\subsection{VP concepts}

Visual Programming (VP) languages enable programming by means of interacting with graphical elements, such as blocks, symbols, and arrows, rather than text. They are known, in general, to promote program comprehension by representing content in two-dimensions (text is considered one-dimensional) and by emphasizing the underlying semantics rather than the syntax \cite{myers90, Shu89}. The contexts where VP is used mostly include learning environments (for both children and novice programmers) and specialized domains \cite{myers90, Nardi93}.

%VP languages have been in existence since GUIs became popular \cite{myers90}. The direct manipulation means afforded by GUIs made these languages possible. Given that touchscreen devices only heighten the direct manipulation interaction style afforded by GUIs, it can be said that VP languages can be as (if not more) conveniently used on touchscreen devices and a somewhat natural fit given that they inherently leverage this interaction style and make minimal or no use of the keyboard \cite{Knaus08, hackett12}. 
VP languages are a somewhat natural fit for use on touchscreen devices because they inherently leverage their interaction style and make minimal or no use of the keyboard \cite{hackett12, Knaus08}. Aspects of VP can be observed even in the syntax-enforcing implementations of \textit{TouchDevelop} \cite{Tillmann11} and the syntax-directed keyboard extension \cite{Almusaly15}. Therefore, incorporating VP concepts in the implementations of even conventional programming languages on touchscreen devices can be useful in advancing their usability. %These devices also enable drawing and manipulation using a stylus which make it possible to realize the diagrammatic programming subset of VP \cite{Shu89,plimmer06_2}. 

%in reference to the nature of interaction. Conversely, these languages or environments have also been said to be programming environments that are a `natural-fit' or best suited for use on touchscreen devices because they inherently leverage the interaction style of these devices as well as make minimal or no use of the keyboard \cite{Knaus08, hackett12}. 

%However, VP is said to use screen space ineffectively in connection with certain tasks requiring the display of more content or more complex program structures \cite{Nardi93}. Therefore, one of the challenges in enabling VP on mobile devices with limited screen space is rendering adequate programming content on them. 

\subsection{Examples of VP-based environments on mobile devices}

%\subsubsection{Educational environments}

Learning environments are one of the main contexts in which visual programming (VP) applications are used. VP environments, similar to and inspired by those commonly used in learning contexts for children and young adults, such as \textit{Scratch} \cite{resnick09} and \textit{Blockly} \cite{fraser13}, have been implemented or adapted for use on touchscreen devices such as smartphones and tablets; examples include \textit{ScratchJr} \cite{Strawhacker15}, \textit{Hopscotch} \cite{fryer14}, \textit{YinYang} \cite{McDirmid11} and \textit{Catroid} \cite{Slany12}. Ihantola et al. \cite{Ihantola13} and Karavirta et al. \cite{Karavirta12} present adaptations of the Parson's puzzles web application \cite{Parsons06} for use on mobile touchscreen devices; these block-based puzzles are designed to help students learn new programming languages.

%Past work enabling VP-based applications on touchscreen devices include papers that emphasize the strong ties between VP and touchscreen interaction \cite{hackett12, hacketttwo, Essl10}, applications for use in educational contexts \cite{Strawhacker15, fryer14, McDirmid11, Slany12, Karavirta12, Ihantola13}, end-user applications \cite{seifert11, danado12}, applications for use in specialized domains such as `music programming' \cite{yang15, salazar14}, and applications enabling specific types of VP, namely, icon-based \cite{bischoff02} and diagrammatic programming \cite{plimmer06_2}. %These papers are briefly discussed below.

%\subsubsection{VP and touchscreen interaction}

Hackett and Cox discuss the natural fit between VP and touchscreen interaction and explore the use of bi-manual interactions on multitouch tablets for a VP environment prototype \cite{hacketttwo, hackett12}. Essl stresses the need to formulate approaches that take into account the input and display capabilities of mobile devices for enabling programming on them and presents an example approach based on data-flow programming for multitouch mobile devices \cite{Essl10}.

%VP-based educational environments for K-12 youth implemented on touchscreen devices such as smartphones and tablets include \textit{ScratchJr} \cite{Strawhacker15}, \textit{Hopscotch} \cite{fryer14}, \textit{YinYang} \cite{McDirmid11}, \textit{Catroid} \cite{Slany12}, \textit{MobileParsons}\cite{Karavirta12}, and an extension of `MobileParsons' by Ihantola et al. \cite{Ihantola13}. Most of these environments are based on a drag-and-drop blocks-based metaphor that reduces syntax and enforces constraints to aid with learning semantics.
%present `MobileParsons', an adaptation of the Parson's puzzles application for use on mobile touchscreen devices. Parson's problems are puzzles that aid in students' learning of programming languages; they are generally implemented as web applications containing statement-blocks which are positioned and ordered using `drag and drop' \cite{Parsons06}. 
%Ihantola et al. \cite{Ihantola13} extend `MobileParsons' by implementing duplicatable statement-blocks with editable elements which allow students to solve a problem in many ways, a feature that is absent in `MobileParsons' where the concrete statement-blocks only allow for a fixed solution. 
%\NOTE{I would shorten the previous paragraph and simply state that many VP based education environments support a blocks based drag and drop programming language.  Maybe even discuss the advantages of such  - spatial reasoning, syntax enforcement, etc.}

%\subsubsection{End-user applications}

Visual programming is also popularly used in building end-user programming applications \cite{Nardi93}. %It has been used in live coding \cite{mclean10} environments for musical applications on mobile multitouch devices; while the music interface developed by Yang and Essl is based purely on VP \cite{yang15}, \textit{miniAudicle} \cite{salazar14} uses a hybrid approach combining textual code entry and VP. 
The \textit{UIBuilder} of \textit{MobiDev} \cite{seifert11}, discussed previously, is also VP-based. %is an end-user environment developed to enable the creation of web-based GUI apps on smartphones \cite{seifert11}. It is a hybrid environment and provides three ways to create the apps: users can use a graphical editor or an image of the UI sketch of the app to build the front-end and type the code for the back-end, or type the entire code for both the front-end and program logic.
\textit{Puzzle} \cite{danado12} is another example of a VP-based end-user programming application built to enable mobile-app development on mobile touchscreen devices.

%\subsubsection{Subsets of VP}

Research focused on specific types of VP include %an `icon-based' programming application developed for robot programming on mobile touchscreen devices \cite{bischoff02} and 
\textit{Freeform} \cite{plimmer06_2}, which is a Visual Basic IDE plug-in based on ``diagrammatic programming" that enables sketching of UI designs using a stylus on touchscreen devices which are recognized and converted to code by the underlying IDE.

\subsection{Discussion}

The key takeaway from this section is that VP elements such as blocks, arrows, and other graphical symbols including the UI components on mobile devices are easily manipulable on mobile touchscreen devices. 
Given that VP is said to use screen space ineffectively when displaying more content or more complex program structures \cite{Nardi93}, one of the challenges in enabling VP on mobile devices with limited screen space is rendering adequate programming content on them. Aspects of VP can be also be incorporated in text-based programming interfaces, i.e. hybrid interfaces, on mobile devices, increasing their usability by facilitating program entry and comprehension.

\section{Theme 4: Multi-device collaboration}

Mobile devices have also been used as auxiliary, internetworked devices in larger programming
environments to provide support for tasks which are inconvenient to achieve using the IDE alone \cite{Parnin10} or to enable new functionality \cite{Bragdon11}. 

%We begin by discussing the motivations for these papers and then cover the specific functionality they enable and the mobile device interactions they involve.

%\subsection{Motivations}

%Professional 

%\subsection{Programming functionality enabled using connected mobile devices}

% also include in discussion that parnin et al paper discusses use of futuristic devices too

\subsection{Augmenting IDEs with mobile devices to distribute programming}

Programmers are frequently faced with tasks such as refactoring and code navigation that require them to view and manipulate multiple programming artifacts simultaneously; using the support provided by IDEs for these tasks can be inconvenient \cite{Parnin10}. Parnin et al. aim to facilitate such tasks by augmenting IDEs with interactive touchscreen devices of varying form factors including portable devices such as tablets, called \textit{CodePads} \cite{Parnin10}.
Snippets of program content from various documents can be visualized concurrently on these devices enabling programmers to view and manipulate artifacts relevant to a particular task in a more convenient fashion without losing focus. 
%The concept is designed mainly for professional programmers to help with both their personal and collaborative tasks. 
%Some of the programming scenarios where \textit{CodePads} can be useful are batch refactoring, navigation through program documents, and managing code histories.

%The auxiliary \textit{CodePads} described by Parnin et al. provide support, external to the IDE, for manipulation and navigation of multiple documents \cite{Parnin10}.
%Parnin et al. envision adding auxiliary interactive touchscreen devices of different form factors called \textit{CodePads} to programming environments (IDEs) to provide better support for tasks requiring simultaneous view and manipulation of multiple programming documents \cite{Parnin10}. 
%Many common programming tasks require programmers to view and change multiple programming artifacts concurrently which can be hard to do using the IDE alone \cite{Parnin10, ko06, parnin06, sillito05}. 
%The features that 

\textit{CodePads} generally display program content, enable pen and multi-touch interactions, provide task-specific functionality, and communicate and collaborate with the primary IDE \cite{Parnin10}. 
Certain \textit{CodePads} are better suited for certain tasks. 
%For example, the authors present \textit{CodePad}-based solutions for a few specific programming tasks which can be hard to perform using the IDE alone. 
%Tablets can be used, for instance, for batch refactoring tasks wherein the relevant code fragments from various documents can be selected in the IDE and sent to the \textit{CodePad}. These code fragments can then be manipulated on the \textit{CodePad} and the changes can be reflected in the IDE.
%\textit{CodePads} can also facilitate navigation tasks that require programmers to keep track of various locations across documents and actions associated with the code in those locations \cite{Parnin10}. %These tasks generally result in a series of quick and haphazard browsing of many documents (a phenomenon called `jitter') when performed in the IDE \cite{singer05, parnin06}. 
%Parnin et al aim to address the limited support IDEs provide for certain programming tasks by adding interactive touchscreen devices called \textit{CodePads} to the programming environments \cite{Parnin10}. 
  %are defined for performing specific programming tasks \cite{Parnin10}. 
The task- and \textit{CodePad}-dependent multi-touch gestures presented in the paper for different task scenarios include %drawing a square for creating a new placeholder class and dragging methods into the class (used in particular refactoring tasks), 
using downward and upward
vertical swipe gestures to select and highlight text, respectively,
two-finger gestures to scroll, merge and split code
fragments, expand and shrink gestures for semantic zoom, 
%a magnifying lens overlay to read a small piece of text without zooming the entire display, tapping on a selected code fragment for it to be focused in the IDE, 
and gestures that communicate
with the IDE. % (for example, to apply the changes made). 
%\textit{CodePads} can minimize ``jitter" by providing visual aids, called ``waypoint markers", of various types for the artifacts visited to help programmers better remember their navigation histories. 
\textit{CodePads} can also be useful for recording code histories, for e.g., by combining a chronological visualization of the programming activity and code annotations added by the programmer using pen input. 

The designs for content display on \textit{CodePads} include using low-fidelity representations
for tasks such as navigation through documents %that require a high-level overview of the artifacts (the actual reads and edits happen in the IDE) 
and using high-fidelity representations for tasks that require manipulation of the program content on the \textit{CodePad}. The representation used can also be dependent on the form factor of the \textit{CodePad}. % and the semantic zoom feature can be used to switch between low and high-fidelity representations. 

%\textit{CodePads} will also have provisions for pen input to enable code annotations helpful in managing code histories.

%The design decisions regarding content display on \textit{CodePads} include using low-fidelity representations for tasks such as \emph{navigation} through documents that require a high-level overview of the artifacts, and using high-fidelity representations for tasks that require \emph{manipulation} of the program content on the \textit{CodePad}, and using the semantic zoom feature to switch between low- and high-fidelity representations. \textit{CodePads} also have provision for pen input to enable code annotations.
%helpful in managing code histories. 

\subsection{Connected mobile devices facilitating access to shared program content during developer meetings}

In the context of developer meetings, the environments generally come equipped with support for a single presenter or displaying a single presentation; developers often switch between different presentations by connecting the projector to different devices, often disrupting task-focus \cite{Bragdon11}. %\textit{Code Space} is an environment developed for facilitating `democratic access' and sharing of program content between developers' personal devices (mobile phones and touchscreen laptops) and a shared display during co-located meetings \cite{Bragdon11}. %The environment uses two Microsoft Kinect sensors in addition to the multitouch shared display and developers' touchscreen devices to enable seamless interactions through hybrid `touch + air' gestures. 
\textit{Code Space} is an environment developed to enable and facilitate some of the common programming tasks performed during developer meetings.  They support code review, editing and grouping code fragments, and bug triage by providing capabilities such as \textit{democratic} access to program content on a shared display, sharing and transfer of program content between developers' devices and between the devices and the display, and annotation \cite{Bragdon11}. 

%The representation of program content on the shared display as well as on the laptops with touchscreen displays is based on the \textit{Code Bubbles} concept \cite{bragdon10} where related code fragments are grouped and various such groupings (known as working sets) are displayed concurrently to aid in program comprehension. A similar representation but with lesser content is displayed on the mobile devices to enable interactions in the \textit{Code Space} environment. 

\textit{Code Space} combines `touch' and `air' gestures to enable interactions between the developers' devices and a multitouch shared display during developer meetings \cite{Bragdon11}. The touch gestures are performed on the developers' mobile touch devices, laptops with touchscreen displays or on the shared display; the hybrid `touch + air' interactions are enabled with two Microsoft Kinect sensors. The \textit{Code Bubbles}-like representations \cite{bragdon10} on the devices are accessed using touch and also provide a mode for annotations. Some of the gestures presented include using a mobile touch device for pointing and manipulating digital objects on the display, bi-manual interactions involving pointing and directional `swipe' gestures to transfer content between a device and a display or between two devices, and vertically holding a mobile device to temporarily display content on the shared display. %, and bi-manual interactions based on \textit{Toolglass} \cite{bier93} to access tool palettes as well as to control gesture modes. 
Typically, in the bi-manual interactions, one hand is used to perform the air gesture (pointing or posture) and the other hand performs the touch gesture on one of the touchscreen devices or display.

A formative study was performed to evaluate \textit{Code Space} with 9 professional developers \cite{Bragdon11}. Overall, the participants reacted positively to the system and most of the `touch + air' pointing gestures were deemed socially acceptable by the participants except the gestures for peer-to-peer transfer where they were required to point at a colleague. 

\subsection{Discussion}

Mobile devices come equipped with a multitude of unique sensors and features which can be leveraged to enable new functionality that are either absent or impossible to implement in conventional programming environments. The challenge lies in exploiting their features and connectedness to explore the opportunities they present for novel applications in programming contexts. 

Connected mobile devices can be especially useful in facilitating collaborative tasks and in distributing programming. Traditional IDEs are generally seen as the domain of individual programmers and contain features that are not conducive to collaborative work \cite{Hardy11}.
Mobile devices can enable ``software peer review" wherein all the team members review the same code on their respective devices and their inputs are made visible to others. A small-scale version of this problem, pair-programming, has been addressed in \textit{CodeGraffiti} \cite{Lichtschlag10} which is discussed in the next section.

Traditional IDEs provide support for a wide variety of programming tasks by incorporating numerous tools but it has been observed that programmers, generally, are either unaware or do not make optimal use of many of these tools \cite{latoza06, shepherd08}. Given that the features included in the IDEs can be cognitively demanding and overwhelming, it can be useful to focus on individual tasks independently by distributing them on peripheral connected mobile devices, also supporting ``separation of concerns" \cite{hursch1995separation}, similar to the approach in \textit{CodePads} \cite{Parnin10}.

%In general, connected mobile devices can be used for any task requiring collective access or access to a subset of program content. This can also result in them enabling new functionality such as being used as a remote control in \textit{Code Space} \cite{Bragdon11}.

\section{Theme 5: Aids for code comprehension}

%In this section, we present past work where support for code comprehension in the form of pen-based annotations \cite{Lichtschlag10, Sutherland13, plimmer06_2, Priest06, Chen07} and sketches linked to code \cite{Baltes14} is provided by the integration of mobile touchscreen devices into the programming environments. 

Programmers exhibit a behavior, both when working independently and during tasks involving collaboration, %such as ``pair-programming",
wherein they often use elements external to the IDE such as paper or whiteboards for taking notes or drawing sketches to complete programming tasks \cite{cherubini07, latoza06, parnin10_2}. 
In this section, we present past work that addresses this behavior and attempts to make these usually transient notes and sketches persistent with the help of mobile touchscreen devices. %linking them to the corresponding code snippets in the IDE for future reference and to increase code comprehension by integrating mobile touchscreen devices into the programming environments. 

\subsection{Pen-based code annotations and linking images to code}

%Programmers exhibit a behavior, both when working independently and during tasks involving collaboration such as ``pair-programming", wherein they often use elements external to the IDE such as paper or whiteboards for taking notes or drawing sketches to complete programming tasks \cite{parnin10_2, cherubini07, latoza06}. These usually transient notes and sketches are made persistent by linking them to the corresponding code snippets in the IDE for future reference and to increase code comprehension . 

Aids for code comprehension include enabling pen-based annotations (the annotations remain attached to the respective code segments despite dynamically changing code) in the IDE \cite{Chen07, Lichtschlag10, Sutherland13, plimmer06_2, Priest06} and linking images (taken using the camera) containing sketches to code in the IDE \cite{Baltes14}. 
Pair-programming, where two programmers collaborate to solve programming tasks,
%- one performs the coding while the other constantly comments or provides insights, 
is addressed in \textit{CodeGraffiti} \cite{Lichtschlag10} by enabling sketching and annotation on a peripheral tablet device used by the non-coder while the same code view is shared on both the coder's computer and the tablet device.  

\subsection{Discussion}

Studies have found that programmers spend more time reading and comprehending code as part of maintenance tasks than they do writing code \cite{Ko05}. %Given that text input using the soft keyboard on mobile devices can be difficult, 
These behaviors can be advantageous for mobile devices provided sufficient aids for program comprehension are incorporated. These aids can range from minor changes in the code representation such as including syntax coloring (or highlighting) and indentation to providing program visualizations such as the call graph. Additionally, tools for browsing from high-level views to low-level details and vice versa, search tools, and context-driven views can be helpful \cite{storey2005theories}. Some of these tools are also included in \textit{CodePads} \cite{Parnin10} and \textit{Code Space} \cite{Bragdon11}. 

Ideas for tools to incorporate on mobile devices to facilitate code comprehension can be drawn from the tools included in IDEs \cite{bassil2001software, storey2005theories} and adapting them appropriately for use on mobile devices. Many of these tools are also programming-language-dependent, such as syntax-coloring. Mobile device features can also be leveraged to provide novel functionality, such as linking camera images to code \cite{Baltes14}.
Augmented reality techniques could be implemented to give collaborating developers a personal window into the source code overlayed with additional comprehension information such as highlighting potential code smells.
%Additionally, mobile devices can be equipped with functionality analogous to Augmented Reality (AR) wherein they can sense or parse a program (displayed external to the mobile device) using its camera, for example, and automatically present visualizations of the code or identify bugs or suggest refactorings. 

%A percentage of the targeted users for programming on mobile devices include end-users and non-expert programmers and hence incorporating code comprehension aids for domain-specific languages may also be important to promote usability \cite{pereira2008program}. 

\section{Theme 6: Gestures for programming tasks}

Touch-based gestures have been formulated for programming tasks such as refactoring, either in an effort to add touch support to traditional IDEs \cite{bacikova15, Biegel14}, or to be used as part of programming environments on mobile touchscreen devices \cite{Raab13}.

\subsection{Gestures formulated based on guidelines}

Biegel et al. \cite{Biegel14} present design decisions for optimizing the GUI of the \textit{Eclipse} IDE for touch access and a set of touch gestures for common refactoring tasks. 
%The changes made to the IDE layout and existing GUI elements were based on the results of a usability study of the default IDE interaction on a touchscreen display. Many factors were considered in the redesigns such as the `fat-finger' problem, handedness of the users, and fatigue experienced by users continuously interacting with a vertical display (also known as `gorilla arm'). As a result, a few GUI elements were enlarged, and most frequently accessed elements were positioned at the bottom and on the left or right side depending on the handedness of the users so that they can access these elements by resting their elbows on the table. The long and linear menus of the \textit{Eclipse} IDE are transformed into relatively bigger radial or italic menu overlays so that the users can easily find and `touch' a desired option. 
%A set of gestures are also presented in the paper to facilitate refactoring tasks. 
These gestures were devised based on the guidelines for mapping gestures to refactorings provided by Murphy-Hill et al. \cite{Hill11}.

An evaluation with 8 participants was done %by Biegel et al. \cite{Biegel14} 
to assess the gestures formulated for refactoring \cite{Biegel14}. 
%In general, the participants found the enhanced GUI unusual to use and while the radial menu needed getting used to, the participants were more familiar with the italic menu since its structure is somewhat similar to the original IDE menu. For the evaluation of the gestures, 
The users were asked to perform certain refactoring tasks using the traditional keyboard and mouse setup and using the formulated gestures both on a touch monitor as well as on a tablet. Most of the users did not remember the shortcuts for the tasks and spent a considerable amount of time (between 5 to 15 seconds) to find the respective commands in the IDE menu in the traditional setup. On the other hand, the users quickly learned and used the gestures on the touch devices 
%(in less than 5 seconds, usually) 
but found that it was hard to select text because of the impreciseness of touch.

\subsection{Touch and pen gestures elicited from users}

In \textit{RefactorPad}, Raab et al. present a gesture set for common editing and refactoring tasks that can be used as part of development environments on touchscreen devices \cite{Raab13}.
The gesture set presented %in \textit{RefactorPad} \cite{Raab13, raab16} 
was obtained from a guessability study similar to that described by Wobbrock et al. \cite{wobbrock09}. The study was designed to elicit both pen and touch gestures from the participants for a list of non-programming-language-specific editing and refactoring tasks which was compiled by studying various editors including \textit{Eclipse}, \textit{Sublime Text}, \textit{Visual Studio} and \textit{Xcode}. The study results showed that more participants preferred to use the pen for performing the gestures and participants often used the same gestures using both touch and pen input for a given task. Multitouch gestures were used for only a few tasks.
%and the participants generally found it easier to perform the `extract method' refactoring than the `inline method'. 
%Popularly used gestures by the participants contained in the recommended set include using the line numbers for selecting lines or a block of code. 
The interaction behaviors of the participants also appeared to be influenced by the mobile operating systems (\textit{Android} or \textit{iOS}) they were most familiar with. 

The programming-specific gestures for touch-enabled IDEs presented by Ba{\'c}{\'\i}kov{\'a} et al. include some of the existing general purpose gestures used in \textit{Android} and \textit{iOS} multitouch platforms, new general purpose gestures enabling certain IDE features such as code folding and semantic zoom, and `drawn' gestures elicited from users for various program constructs such as class declaration and loop statements \cite{bacikova15}. 

\subsection{Discussion}

Gestures are one of the primary interaction techniques used on mobile devices. The programming tasks mentioned above are normally performed in IDEs using either keyboard shortcuts, typing, selecting items from long menus, or a combination of these. While keyboard shortcuts and menu-based approaches are far from intuitive, gestures have the potential to be more intuitive, guessable, and memorable \cite{Biegel14, kristensson2007command, Norman10}.  

Gestures can be devised, for example, to perform tasks such as refactoring or to enter code or enable different functionality depending on the context. However, they will have to be designed carefully and standardized over time to be ``usable" \cite{Norman10}. Elicitation studies, such as the ones mentioned above, provide a good method to define gesture-sets that are reflective of user behaviors in a particular context \cite{wobbrock09}. Based on the observations in the above studies, it may be good practice to design gestures that are consistent with the mobile operating systems on which they will be used \cite{bacikova15, Raab13}.
Additionally, pen-based gestures can be used for text-manipulation tasks because of their greater precision and ease of use \cite{Raab13}.

\vspace{3pt}

\section{Summary and Future Directions}

The utility of mobile touchscreen devices in programming situations has not been regarded with much enthusiasm. We have not only reviewed promising uses for these devices in programming, but also hinted at the space of unexplored capabilities of these devices in these contexts. Their ubiquity and features allow them to favorably permeate contexts ranging from solo programmers' workspaces to larger interactive programming environments.  

Mobile devices can, for example, be equipped with functionality that uses the camera to sense the source code displayed external to the mobile device (e.g. on a white board) and automatically present related content, such as last commit or who's working on it now.  It can suggest refactorings, or highlight potential code smells - all on the personal mobile device of one developer without disturbing the shared view.
%For example, the AR-based ideas discussed earlier can be expanded to many connected devices with each providing different functionality. 
Location sensing on the devices can be used to detect if users are travelling, enable access to their code and automatically synchronize changes with the desktop version.

%see wigdor book page 10 for touchscreen usability and that it helps leverage spatial memory; and copying wimp is not a solution.
% copied from 'collaborative code reviews' paper: Since natural and multimodal interaction can be particularly well suited for tasks with high mental demands (Oviatt et al. 2004), problem solving, and collaboration (Shaer and Hornecker 2009), leveraging the capabilities of interactive surfaces for collaborative code inspection tasks seems natural.

%We present below the key observations made and suggest future directions for research.

While reinventing the wheel is not necessary to enable usable programming on mobile devices and a number of existing practices and designs from conventional programming environments can be used to incorporate programming functionality on mobile devices, these practices and designs will have to be adapted to suit the interaction style of mobile devices. Mobile devices come equipped with many features that allow for numerous designs wherein gestures can replace operations commonly performed with the mouse, multimodal input can be used for program entry and editing, and the GUI elements can be visualized in various ways. %Additionally, mobile devices present opportunities for new applications when embedded in larger programming environments.  

%One of our main observations 
%in the work discussed 
%is that the programming support enabled and the (touchscreen) interaction design aspects are two sides of the same coin - they may seem different but are interconnected topics. Therefore, we propose that, using these devices for programming requires rethinking the conventional methods of programming in the light of the interaction capabilities afforded by these devices.

%We present the past work done under six main themes. %, namely, `For smartphones, on smartphones', `Programming language-driven interface designs', `Visual programming-based applications', `IoT and Programming', `Aids for code comprehension', and `Gestures for programming tasks'. 
The six themes presented in the survey are not independent and enabling usable programming conforming to any theme would require borrowing ideas from multiple themes. %We provide examples of this interconnectedness below. 
For example, %we identify 
mobile app %development as one of the main areas for which mobile devices will be used for programming. These 
development environments can benefit from existing programming language designs adapted to enable convenient code entry and manipulation on the devices. %Existing programming languages can be adapted for mobile device use by, for example, incorporating visual programming-based elements (including commonly used mobile UI components) in their designs and gestures for inputting code or performing programming tasks. 
Connected mobile devices can be used for collaborative tasks such as code review as described in the multi-device collaboration theme. This requires enabling task-specific functionality on the devices and aids for code comprehension such as search and navigation tools and code visualizations that are specific to the task. These aids may also be programming-language-dependent and use the language's syntax and semantics to provide the necessary visualizations. 

%Since the programming possibilities are numerous and involve many variables such as device types, contexts of use, and target users, the themes presented in this survey can also serve as good starting points to approach the problem. For example, identifying particular `contexts' of use can narrow down the choice of devices and/or programming capabilities supported. 
%The themes presented in this survey can also serve as starting points to approach the problem of enabling programming functionality on the mobile devices. For example, we have discussed work where mobile devices are used as peripheral devices in larger contexts to enable some programming capability. While some of them use mobile devices to augment IDEs to enable functionality lacking in the IDEs \cite{Parnin10, Lichtschlag10}, some others aim to integrate these devices into environments to enable new functionality \cite{Bragdon11}. 
There are numerous studies on the behaviors of programmers as well as the functionality and features of conventional IDEs \cite{ko06, latoza11, latoza06, parnin10_2}. These studies can serve as useful starting points to identify if any shortcomings in the conventional programming setups can be mitigated by integrating mobile devices or if mobile devices can enable novel interaction or programming capabilities in these environments.

There are also many research topics that need to be explored in reference to the programming languages that can be supported on mobile devices. It is known that traditional IDEs have been designed keeping in mind the interaction style (WIMP) of the mouse and keyboard \cite{Raab13} but it is unclear whether conventional programming language designs also have strong affinities with the interaction style of desktops. The extent of these affinities may play a role in determining their usability and adaptation for use on mobile devices. %need citations to include here

In summary, we have presented the diverse body of research with respect to using mobile touchscreen devices for programming, identified the main themes of contributions, and presented the programming capabilities supported, the interaction capabilities of the devices leveraged, and users' acceptance of the implementations in each theme. We have identified key challenges and presented potential directions for future work.  As researchers move forward in exploring the use of mobile devices in programming contexts, we hope this survey can serve as a starting point to understand what has been done and a roadmap for areas that should be further explored.

\bibliographystyle{acm-sigchi}
\bibliography{sample}

%table

%final table starts here

%\setlength{\arrayrulewidth}{1mm}
%\setlength{\tabcolsep}{18pt} %alters only left and right spacing in column text

\renewcommand{\arraystretch}{2} %alters only up and down spacing in rows

\begin{landscape}% Landscape page
\begin{table}
%\medskip\noindent

    %\begin{center} % Center table
   % {\tabulinesep=1.5mm
        \resizebox*{\linewidth}{\textheight}{ \begin{tabular}  { |c |c !{\VRule[2.5pt]} c| c !{\VRule[2.5pt]} c| c |c| c| c !{\VRule[2.5pt]} c| c| c |c !{\VRule[2.5pt]} c |c| c |c| c|c |c |c !{\VRule[2.5pt]} c |c |c|  c !{\VRule[2.5pt]} c |c |c !{\VRule[2.5pt]} c  c  c c  c  c c |} %35 columns
  %   \resizebox*{\linewidth}{\textheight}{ \begin{tabu} to 1\linewidth { |X[c]| X[c] | X[c]| X[c] |X[c] |X[c]| X[c]| X[c]| X[c] |X[c]| X[c]| X[c]| X[c]| X[c]| X[c] |X[c] |X[c] |X[c]| X[c] |X[c]| X[c] |X[c] |X[c] |X[c] |X[c] |X[c]|  X[c] |X[c] |X[c] |X[c] |X[c]  X[c]  X[c] X[c]  X[c]  X[c] X[c] X[c] X[c]|} % creating 39 columns

                \hline 
               
  \multicolumn{2}{|c|}{\parbox{3em}{Mobile device types}}
  & \multicolumn{2}{l|}{\parbox{3em}{Role of mobile devices}} & \multicolumn{5}{c|}{\parbox{6em}{Programming Language}} & \multicolumn{4}{c|}{\parbox{4.5em}{Type of implementation or prototype}} & \multicolumn{8}{c|}{Programming support} & \multicolumn{4}{c|}{\parbox{5em}{Interaction types}} & \multicolumn{3}{c|}{\parbox{3em}{User studies}} & & & & & & &  \\  \cline{1-28}
  & & & & & & & & & & & & & & & & & &
  & & & & & & & & & & & & & & & &  \\
  
  %\hline
   \rot{Smartphones} & \rot{Tablets} & \rot{Primary} & \rot{Auxiliary}  & \rot{Imperative programming} & \rot{Concatenative programming} & \rot{New programming language} & \rot{Visual or hybrid programming} & \rot{Not programming language specific} & \rot{Standalone programming environment} & \rot{Mobile app development-based} & \rot{Web-based} & \rot{IDE-based} & \rot{Structured editing}& \rot{Autocompletion} & \rot{Code Annotation} & \rot{Refactoring tasks} & \rot{Collaborative tasks (e.g., code review and pair-programming)} & \rot{Debugging} & \rot{Visual representations for code comprehension} & \rot{Software build tools} & \rot{Touch (including multi-touch and gestures)} & \rot{Touch + Pen} &  \rot{Touch + Speech} & \rot{Touch + camera}  & \rot{Elicitation study} & \rot{Usability study} & \rot{Longitudinal study} & & & & & & &  \\

\hline
\hline
 %    \multirow{6}{*}{Touchscreen devices as primary platforms for text-based programming applications}
  
  %\parbox[t]{2mm}{\multirow{6}{*}{\rotatebox[origin=c]{90}{Touchscreen devices  as primary platforms for text-based programming applications}}}

%learning environments

x& &x & & & &x & & &x &x & & & x&x& & & & & &x &x & & & & & & x&  \multicolumn{6}{l|}{TouchDevelop \cite{Tillmann11}} &   \multirow{3}{7em}{For Smartphones, On Smartphones}   \\ \cline{1-34}

x& &x & &x & & &x & &x & x&x & & & & && & & &x & & & & x& &x & &  \multicolumn{6}{l|}{Mobidev \cite{seifert11} } &   \\ \cline{1-34}

x& & x& &x & & & & &x &x & & & & && & &x &x &x &x & & & & & x& &  \multicolumn{6}{l|}{GROPG \cite{Nguyen13} } &   \\ \cline{1-34}

\hline \hline

& x &x & & & x& & & &x & & & & & && & &x &x & &x & & & & & & &  \multicolumn{6}{l|}{Touching Factor \cite{Hesenius12}} &   \multirow{4}{7em}{Programming-Language-Driven Interface Designs}  \\ \cline{1-34}

x&x &x & &x & & & & &x & & & &x && & & & &x & & & &x & & & & &  \multicolumn{6}{l|}{Deverywhere \cite{Feldman15} }   & \\ \cline{1-34}

&x &x & &x & & & & & & & & &x & x& & & & & & &x & & & & & x& &  \multicolumn{6}{l|}{\parbox{10em}{Syntax-directed Keyboard extension \cite{Almusaly15} } }  &  \\ \cline{1-34}

x& &x & &x & & & & &x & & & &x & & & & & x& &x &x & & & & & x& &  \multicolumn{6}{l|}{\parbox{10em}{Static scaffolding methods for Java programming \cite{mbogo2016design} } }  &  \\ \cline{1-34}

\hline \hline

x&x &x & & & & &x & &x & & & &x& & & & & & &x &x & & & & & x& &  \multicolumn{6}{l|}{\parbox{10em}{ScratchJr \cite{Strawhacker15}, Hopscotch \cite{fryer14}, Catroid \cite{Slany12}, YinYang \cite{McDirmid11}} }   & \multirow{2}{7em}{Visual-Programming-Based Applications (examples)} \\ \cline{1-34}

x&x &x & & & & & x& & & &x & &x& & & & & x& &x & x& & & & &x & &  \multicolumn{6}{l|}{\parbox{10em}{Mobile Parsons \cite{Karavirta12, Ihantola13} } }   & \\ \cline{1-34}

%x&x &x & & & & & x& & & &x & &x& & & & & x& & & x& & & & &x & &  \multicolumn{6}{l|}{\parbox{10em}{Visual programming and touchscreen interaction papers \cite{hacketttwo, hackett12, Essl10} } }   & \\ \cline{1-34}

%x&x &x & & & & & x& & & &x & &x& & & & & x& & & x& & & & &x & &  \multicolumn{6}{l|}{\parbox{10em}{End-user applications \cite{yang15, salazar14, seifert11, danado12} } }   & \\ \cline{1-34}

%x&x &x & & & & & x& & & &x & &x& & & & & x& & & x& & & & &x & &  \multicolumn{6}{l|}{\parbox{10em}{Subsets of visual programming \cite{bischoff02, plimmer06_2} } }   & \\ \cline{1-34}

\hline \hline

&x & &x & & & & & x& & & &x& & &x &x &x &x &x & & &x & & & & & &  \multicolumn{6}{l|}{CodePads \cite{Parnin10}} &   \multirow{2}{7em}{Multi-device Collaboration}  \\ \cline{1-34}

x& x &x & & & & & &x & & & &x && &x & &x &x &x &x & &x & & & & x& &  \multicolumn{6}{l|}{\parbox{10em}{Code Space \cite{Bragdon11}} }   & \\ \cline{1-34}

\hline \hline

& x& & x& & & & &x & & & &x && &x & & & & & & & x& & & &x & &  \multicolumn{6}{l|}{\parbox{10em}{Code annotation papers \cite{Sutherland13, Priest06, Chen07}} }   & \multirow{3}{7em}{Aids for Code Comprehension} \\ \cline{1-34}

x&x & &x & & & & &x & & & &x & &&x & &x &x & & & &x & & & & & &  \multicolumn{6}{l|}{\parbox{10em}{CodeGraffiti \cite{Lichtschlag10}} }   & \\ \cline{1-34}

x& x& &x &x & & & & & & &x &x & && & & & & & & & & &x & & & &  \multicolumn{6}{l|}{\parbox{10em}{SketchLink \cite{Baltes14}} }   & \\ \cline{1-34}

\hline \hline

 &x &x & & & & & &x & & & & x& && &x & & & & &x & & & & &x & & \multicolumn{6}{l|}{\parbox{10em}{Touchifying an IDE \cite{Biegel14}} } &  \multirow{3}{7em}{Gestures for Programming Tasks} \\ \cline{1-34}

&x &x & & & & & &x & & & & x& & & && & & & x& x& & & &x &x & &  \multicolumn{6}{l|}{\parbox{10em}{Programming-specific gestures \cite{bacikova15} } } &   \\ \cline{1-34}

&x &x & & & & & & x& x& & &x && & &x & & & & & & x& & & x& & &  \multicolumn{6}{l|}{RefactorPad \cite{Raab13} }  &  \\ \cline{1-34}

        \hline
        \end{tabular} }
       %\end{tabu} } }
        \caption{Summary of the survey listing the papers that use mobile touchscreen devices for programming and the themes they fall under (rows) and the key attributes of their contributions (columns)}
        \label{table:1}

        \end{table}
        
    %\end{center}
\end{landscape}

\end{document}